\documentclass[default]{sn-jnl}

\jyear{2021}%

\theoremstyle{thmstyleone}%
\theoremstyle{thmstyletwo}%
\newcommand{\rsun}{$R_{\odot}$}
\usepackage{hyperref}
\usepackage{amsmath}
\usepackage{amstext}
\usepackage{graphicx}
\usepackage{xcolor}
\usepackage{comment}
\usepackage{caption}
\usepackage{subcaption}
\usepackage{makecell}
\usepackage{mathtools}
\theoremstyle{thmstylethree}%
\usepackage{natbib}

\begin{document}

\title[Calibration of VELC detectors on-board Aditya-L1 mission]{Calibration of VELC detectors on-board Aditya-L1 mission}

\author[1]{\fnm{Shalabh} \sur{Mishra}}
\author*[1]{\fnm{K.} \sur{Sasikumar Raja}}\email{sasikumar.raja@iiap.res.in}
\author[1]{\fnm{Sanal Krishnan} \sur{V U}}
\author[1]{\fnm{Venkata Suresh} \sur{Narra}}
\author[1]{\fnm{Bhavana Hegde} \sur{S}}
\author[1]{\fnm{Utkarsha} \sur{D.}}
\author[1]{\fnm{Muthu Priyal} \sur{V}}
\author[1]{\fnm{Pawan} \sur{Kumar S}}
\author[1]{\fnm{Natarajan} \sur{V}}
\author[1]{\fnm{Raghavendra Prasad} \sur{B}}
\author[1]{\fnm{Jagdev} \sur{Singh}}
\author[1]{\fnm{Umesh Kamath} \sur{P}}
\author[1]{\fnm{Kathiravan} \sur{S}}
\author[1]{\fnm{Vishnu} \sur{T}}
\author[1]{\fnm{Suresha} \sur{}}
\author[1]{\fnm{Savarimuthu} \sur{P}}
\author[2]{\fnm{Jalshri H} \sur{Desai}}
\author[2]{\fnm{Rajiv} \sur{Kumaran}}
\author[3]{\fnm{Shiv} \sur{Sagar}}
\author[2]{\fnm{Sumit} \sur{Kumar}}
\author[2]{\fnm{Inderjeet Singh} \sur{Bamrah}}
\author[1]{\fnm{Amit} \sur{Kumar}}

\affil*[1]{\orgname{Indian Institute of Astrophysics}, \orgaddress{\street{2nd Block, Koramangala}, \city{Bangalore}, \postcode{560034}, \state{Karnataka}, \country{India}}}

\affil[2]{\orgdiv{Space Applications Centre}, \orgname{ISRO}, \orgaddress{\street{Satellite Road}, \city{Ahmedabad}, \postcode{380015}, \state{Gujarat}, \country{India}}}

\affil[3]{\orgdiv{U R Rao Satellite Centre}, \orgname{ISRO}, \orgaddress{\street{HAL Old Airport Road}, \city{Bengaluru}, \postcode{560017}, \state{Karnataka}, \country{India}}}


\abstract{
Aditya-L1 is the first Indian space mission to explore the Sun and solar atmosphere with seven multi-wavelength payloads, with Visible Emission Line Coronagraph (VELC) being the prime payload. It is an internally occulted coronagraph with four channels to image the Sun at 5000 \AA~ in the field of view 1.05 - 3 \rsun, and to pursue spectroscopy at 5303 \AA, 7892 \AA~ and 10747 \AA~ channels in the FOV (1.05 - 1.5 \rsun). In addition, spectropolarimetry is planned at 10747 \AA~ channel. Therefore, VELC has three sCMOS detectors and one InGaAs detector. In this article, we aim to describe the technical details and specifications of the detectors achieved by way of thermo-vacuum calibration at the CREST campus of the Indian Institute of Astrophysics, Bangalore, India. Furthermore, we report the estimated conversion gain, full-well capacity, and readout noise at different temperatures. Based on the numbers, it is thus concluded that it is essential to operate the sCMOS detectors and InGaAs detectors at $-5^{\circ}$ and $-17^{\circ}$ C, respectively, at the spacecraft level.
}




\maketitle

\section{Introduction}\label{sec1}

ADITYA-L1 is the first Indian solar space mission that is expected to provide valuable insights on solar dynamics with seven payloads on board \citep{Seetha2017}. Visible Emission Line Coronagraph (VELC), an internally occulted coronagraph, is the prime payload of the mission with imaging, spectroscopy, and spectropolarimetry channels being concomitantly operational for observation close to the solar limb \citep{BRP2017, Singh2019}. The configuration of VELC is to enable the payload in order to image solar corona at 5000\AA~  from 1.05 \rsun~ to 3 \rsun~ (\rsun~ represents the solar radii) (i.e., $0.28^{\circ}$ - $0.8^{\circ}$) with a plate scale of 2.5\"/pixel \citep{BRP2017}.

The design configuration of VELC facilitates synchronous spectroscopic observations at three solar emission lines i.e., Fe XIV (at 5303\AA), Fe XI (at 7892\AA) and Fe XIII (at 10747\AA) with a distinct spectral resolution of 28, 31 and 202 m\AA/pixel respectively, over the field of view (FOV) of 1.05\rsun~ to 1.5\rsun~ (i.e., $0.28^{\circ}$ to $0.4^{\circ}$) \citep{BRP2017}. The Fe XIII (at 10747\AA) channel also facilitates to study coronal magnetic fields using dual-beam spectropolarimetry.

The observations in the solar imaging channel at 5000\AA~ alongside spectroscopic observations at 5303\AA~ and 7892\AA~ are accomplished using an sCMOS detector with good spectral response in a range of 4000-7000\AA. The spectroscopy and spectro-polarimetry observations at 10747\AA~ are accomplished using an InGaAs detector with spectral response in the range of 9000-17000\AA. The complex thermal management of the payload ensures that the stringent requirement of chip temperatures at each detector is achieved. The functional control of parameters such as bias, temperature, etc., are augmented to achieve the best in-orbit performance of the payload. 

The detector electronics comprises primarily three functional modules, namely, the Detector Proximity Electronics (DPE), Digital Control and Processing Electronics (DCPE), and Payload Power Electronics (PPE). The DPE comprises the detector, its bias, and the data and clock interface. DCPE receives the clock and transmits the serialized data. It is also responsible for generating control signals and data processing, including onboard intelligence for CME detection \citep{Ritesh2018}. PPE addresses the power requirements of the modules above. The detector electronics interfaces with Onboard computer (OBC) and Onboard Data Handling and Storage (ODHS) unit for payload functionality including commanding and data acquisition and onboard data storage.

\section{Technical details of the detectors}\label{sec2}

The Detector Head Assembly (DHA) incorporates the area array imaging sensor, mechanical interface of DHA for precision mounting on the optical bench of VELC, cold finger thermal interface for optimal temperature control and DPE. The perspicacity to perceive the science objectives of the payload and accordingly determine the optimal imaging sensor to carry out the operations.

\subsection{Scientific grade CMOS sensor}\label{subsec2}

High-speed imaging, high resolution, and exceptionally low readout noise enable the utilization of scientific grade CMOS CIS2051 sensor for VELC payload in imaging and two spectroscopy channels, i.e. 5303\AA~ and 7892\AA. The sensor comprises an array of 2160(H)$\times$2560(V) photodiodes with a pixel dimension of 6.5µm$\times$6.5µm. The sensor is divided into two independent sections (1080$\times$2560 top and bottom card) with independent controls. The sensor is equipped with dual column-level amplifiers so that it can be configured for either 1X or 2X analog gains in Low Gain mode and 10X or 30X analog gains in High Gain mode. The output of both the amplifier levels is concurrently provided to dual-level 11-bit ADCs. Considering the design framework, at any juncture, 22-bit pixel data corresponding to both the gains (11-bit each) for each half of the sensor is available for readout. 

The vital specifications of the sCMOS detector electronics are detailed in Table \ref{CMOS}.

\begin{table}[!ht]
\begin{center}
\caption{Specifications of sCMOS detector}\label{CMOS}%
\begin{tabular}{@{}llll@{}}
\toprule
S. No. & Parameter  & Description \\
\midrule
1.    & Sensor Format   & 2160 $\times$ 2560  \\
2.    & Pixel Size   & $ 6.5\mu m \times 6.5\mu m$ \\
3.    & Spectral Range   & 4000 \AA~ to 9000 \AA  \\
4.    & Readout Mode   & Rolling Shutter  \\
5.    & Number of ports   & \makecell {Two (22-bit) \\ (11-bit low gain, 11-bit high gain)}  \\
6.    & Gain   & Four (1X, 2X, 10X, 30X)  \\
7.    & Integration time   & 10ms to 100sec  \\
8.    & Detector configuration mode   & JTAG  \\
9.    & Operating Mode   & Read then integration  \\
10.   & Pixel Readout (MHz)   & 52.5  \\
11.   & Data Rate (Mbps)   & 52.5 parallel  \\
12.   & Payload data interface   & SerDes  \\

\botrule
\end{tabular}
\end{center}
\end{table}

\subsection{InGaAs FPA sensor}\label{subsec3}

Accommodating the sensor non-uniformity, dark current, full-well capacity, and longer integration times, the InGaAs 640$\times$512 focal plane photodiode array from Chungwa is utilized in the DHA fabrication for the spectro-polarimetry channel of VELC. The sensor is responsive in the SWIR band, i.e., 9000\AA to 17000\AA. The sensor is fabricated such that the photodiode array is attached to the Readout Integrated Circuit (ROIC) utilizing indium bumps.

Considering the notable integration times, appreciable in comparison to the readout time of the detector, integrate while read (IWR) methodology of operation is preferred. The detector output is read out using a single port, which provides an analog output. The output is digitized (12-bit) in FECE (Front-End Camera Electronics) and provided to DCPE through 3V CMOS interface for further processing. The vital specifications of the IR detector electronics are detailed in Table \ref{IR}.

\begin{table}[!ht]
\begin{center}
\caption{Specifications of IR detector}\label{IR}%
\begin{tabular}{@{}llll@{}}
\toprule
S. No. & Parameter  & Description \\
\midrule
1.    & Sensor Format   & 512 $\times$ 640  \\
2.    & Pixel Size   & $ 25\mu m \times 25\mu m$ \\
3.    & Spectral Range   & 9000 \AA~ to 17000 \AA  \\
4.    & Readout Mode   & Rolling Shutter  \\
5.    & Number of ports   & One  \\
6.    & Gain   & Two (LG, HG)  \\
7.    & Integration time   & 50ms to 100sec  \\
8.    & Detector configuration mode   & Serial  \\
9.    & Operating Mode   & Read while integration  \\
10.   & Pixel Readout (MHz)   & 6.56  \\
11.   & Data Rate (Mbps)   & 26.25 serial  \\
12.   & Payload data interface   & 3V LVDS  \\

\botrule
\end{tabular}
\end{center}
\end{table}

\section{Thermo vacuum-chamber and calibration setup}\label{sec3}

VELC has a complex thermal management system considering all the different subsystems and a stable operational temperature at the sensor chip is paramount. During the stand-alone testing of the detectors, the desired temperature was obtained by connecting one end of the mini-heat pipe to detector head assembly and the other end to the external Thermal Control Unit (TCU) via heat exchanger for the thermal flow as shown in Figure-1. Note that after integration of the payload, TCU was replaced by a radiator plate. The details on the coupling of the detectors and the radiator plate will be presented elsewhere.

\begin{figure}[!ht]
  \centering
  \begin{subfigure}[b]{0.45\linewidth}
    \includegraphics[width=1.0\linewidth]{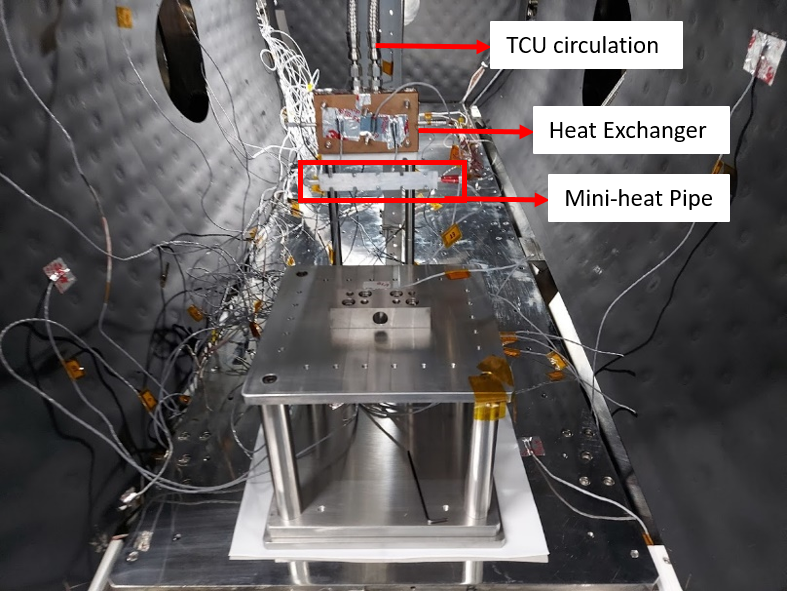}
    \caption{}
  \end{subfigure}
  \begin{subfigure}[b]{0.45\linewidth}
    \includegraphics[width=1.0\linewidth]{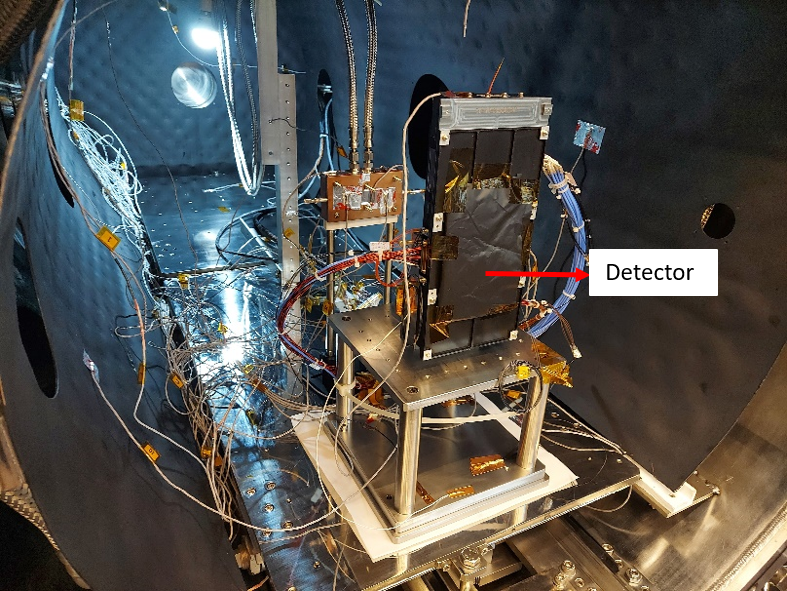}
    \caption{}
  \end{subfigure}
  \caption{Subplot (a) shows the cooling plate interface calibration unit and Subplot (b) shows the DHA after mounting in Thermo Vaccum Chamber for calibration at CREST campus of the Indian Institute of Astrophysics.}
\label{fig:DHAcalib}
\end{figure}

Considering the limitation in mounting a temperature sensor at the detector chip, the closest place for monitoring is at the cooling plate interface. Therefore, a calibration setup, as shown in Figure \ref{fig:DHAcalib}, is designed to estimate the temperature gradient between the detector chip and cooling plate interface by simulating the inherent detector power dissipation using a heater. A sensor is placed at the detector chip interface and another at the cooling plate interface. The test was carried out to measure all the desired chip temperature levels for both sCMOS (visible) and InGaAs (IR) detectors.

Considering the science objectives of the payload along with the system limitations in terms of scatter and efficiency, the dark response of the DHAs, along with performance at various integration times and temperature, becomes of paramount importance \cite{Postma2011}. The DHA calibration inside the thermo-vacuum chamber (TVAC), as shown in Figure \ref{fig:DHAcalib}, was carried out at various temperature levels in order to determine the absolute operating temperature for the respective DHAs to corroborate optimum system performance. The calibration was carried out at $10^{-5}$ mbar vacuum with a thermal control unit established to provide a temperature stability of $0.5^{\circ}$ C. 

\section{Results and Discussions}\label{sec4}

\begin{figure}[!ht]
  \centering
  \begin{subfigure}[b]{0.4\linewidth}
    \includegraphics[width=\linewidth]{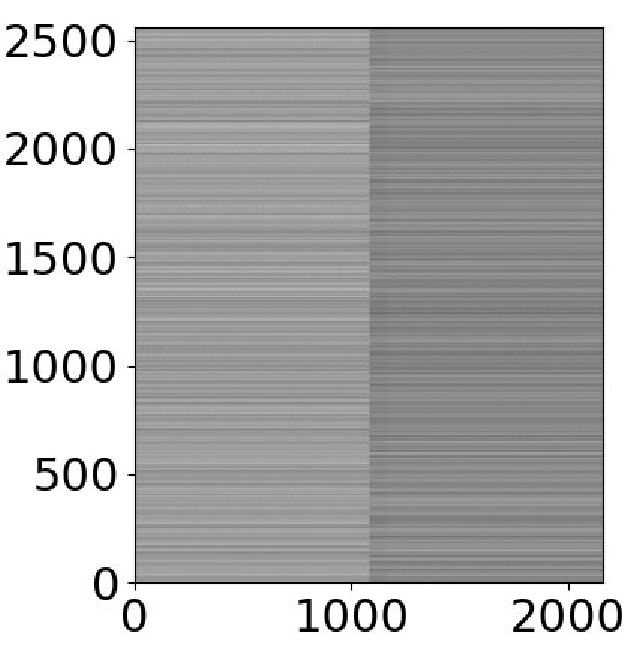}
    \caption{}
  \end{subfigure}
  \begin{subfigure}[b]{0.4\linewidth}
    \includegraphics[width=\linewidth]{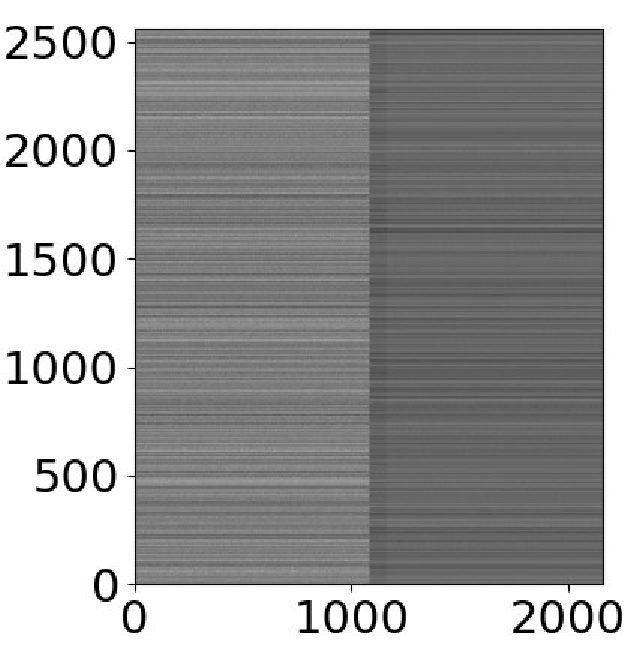}
    \caption{}
  \end{subfigure}
  \caption{Subplot (a) shows the combined low gain image of both top and bottom card of sCOMS detector and subplot (b) shows the combined high gain image of both top and bottom card of the sCMOS detector}
  \label{fig:Dark_CMOS}
\end{figure}

\begin{figure}[!ht]
  \centering
  \begin{subfigure}[b]{0.4\linewidth}
    \includegraphics[width=\linewidth]{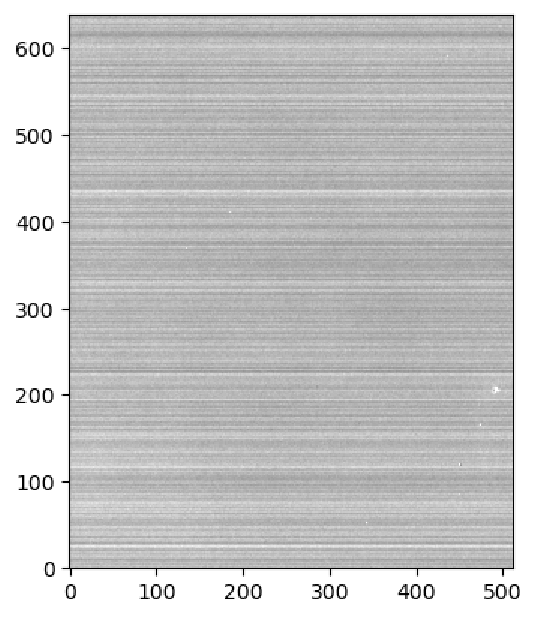}
    \caption{}
  \end{subfigure}
  \begin{subfigure}[b]{0.4\linewidth}
    \includegraphics[width=\linewidth]{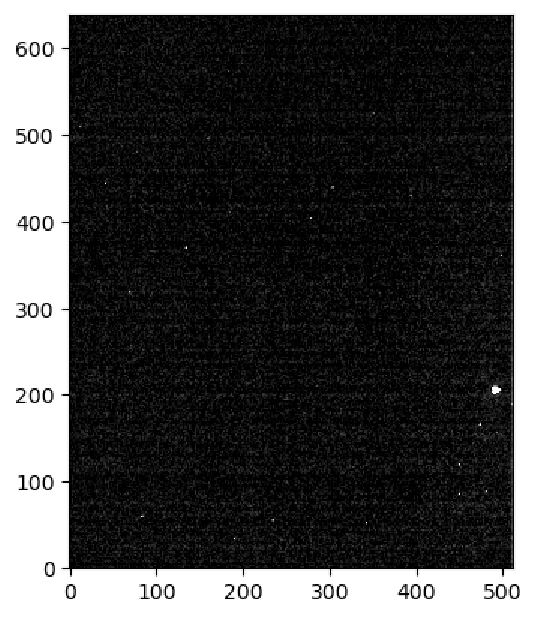}
    \caption{}
  \end{subfigure}
\caption{Subplots (a) and (b) shows the low gain and high gain images of InGaAs detector respectively}  
\label{fig:Dark_IR}
\end{figure}

Set of dark frames of integration time that range from 10ms to 100s is obtained for visible detectors and 50ms to 100s is obtained for IR detector. Furthermore, for flat-fielding, light exposures are acquired with different exposure times such that $\approx 80-90\%$ of the full well capacity is attained. Since the readout for visible and IR detectors is 11-bit and 12-bit respectively, the integration time varies based on the specified gain (during light exposures) and by illuminating the constant flux on the detectors. The dark images obtained using visible and IR detectors is as shown in Figure \ref{fig:Dark_CMOS} and Figure \ref{fig:Dark_IR}, respectively.

\subsection{Estimation of mean and standard deviation of the images}

If `i' and `j' represent the pixel coordinates and `z' represents the frame number, the image $I_z(\,i,j)\,$ represents the intensity values of $z^{th}$ frame at $(\,i,j)\,^{th}$ pixel coordinates. To measure the mean dark counts and standard deviation of the images, if the integration time is greater than 20s, we acquired eight frames, and if the integration time is less than or equal to 20s, we acquired 16 frames per set. 

\subsubsection{Mean calculation}\label{sec:mean}
First, pixel-to-pixel mean of all frames is calculated over the entire frame acquired using the relation
\begin{equation}\label{eq:one}
  I(i,j) = \frac{1}{n}\sum_{z=1}^{n} I_z(\,i,j)\, 
\end{equation}

where \textit{n} is the number of frames acquired and $I(\,i,j)\,$ is the image of each pixel averaged over all the frames. Then mean of $I(\,i,j)\,$ t is calculated using, 
\begin{equation}\label{eq:two}
  Mean = \frac{1}{N\times M}\sum\limits_{i=1}^N\sum\limits_{j=1}^M I(\,i,j)\,
\end{equation}

where \textit{N} and \textit{M} represents the pixel coordinates.

\subsubsection{Standard deviation calculation}
First, pixel-to-pixel standard deviation of all frames is calculated over the entire frame acquired using the relation
\begin{equation}\label{eq:three}
    \sigma (\,i,j)\, = \sqrt{\frac{1}{n-1} \sum\limits_{z=1}^n (I_z(\,i,j)\, - \overline{I(\,i,j)\,})^2},
\end{equation}
where $\sigma (\,i,j)\,$ represents the standard deviation matrix and $\overline{I(\,i,j)\,}$ represents the mean of the $I(\,i,j)\,$ as discussed in Section \ref{sec:mean}.

Furthermore, the standard deviation is obtained by averaging the acquired $\sigma (\,i,j)\,$ matrix using, 
\begin{equation}
    SD = \frac{1}{N\times M}\sum\limits_{i=1}^N\sum\limits_{j=1}^M \sigma(\,i,j)\,
\end{equation}

\begin{figure}[!ht]
    \centering
    \begin{subfigure}{1.0\textwidth}
        \includegraphics[width=\textwidth]{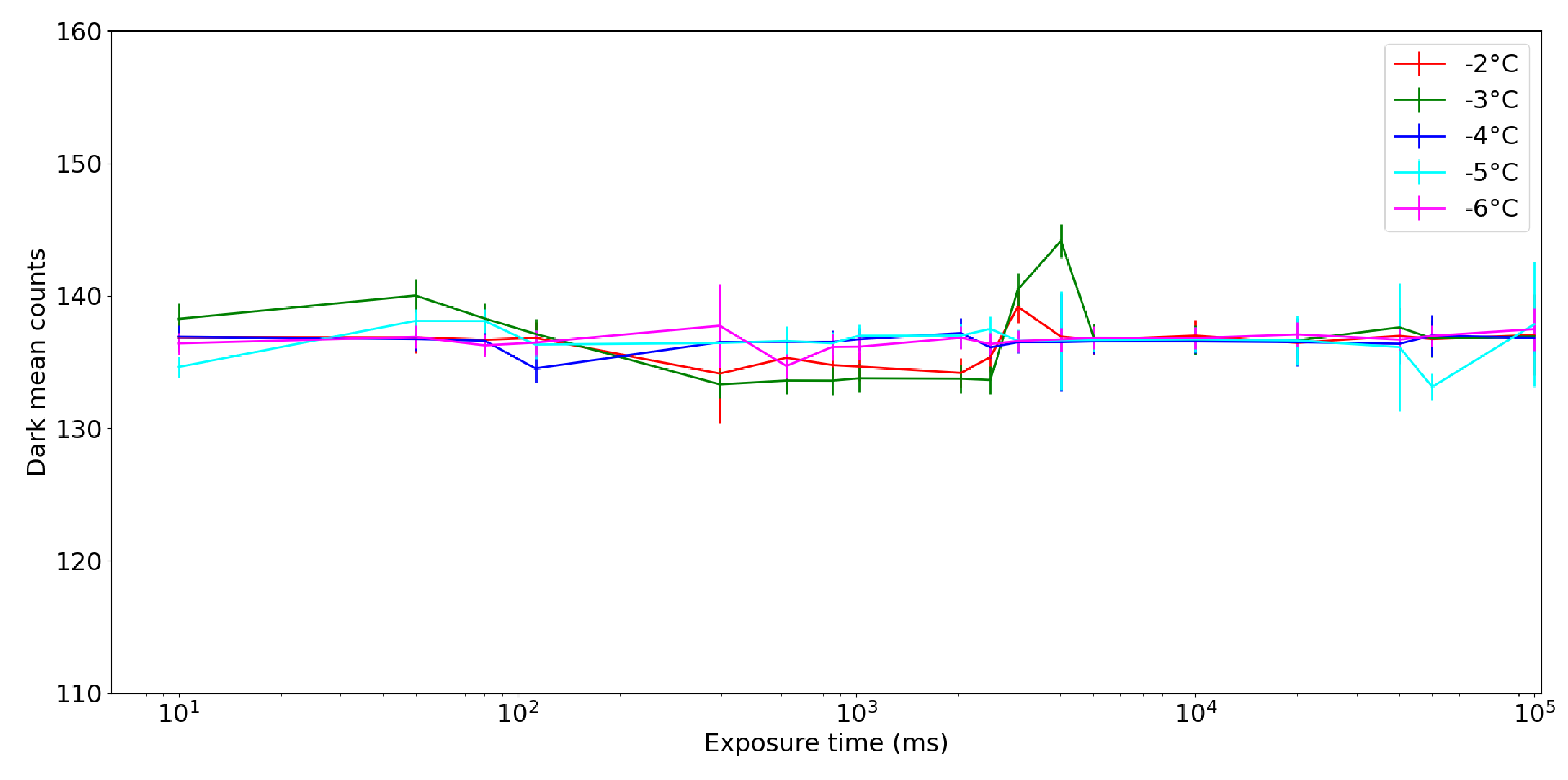}
        \caption{Mean dark variation for sCMOS with deviation as error bar}
        \label{fig:dark30}
    \end{subfigure}
    
    \begin{subfigure}{1.0\textwidth}
        \includegraphics[width=\textwidth]{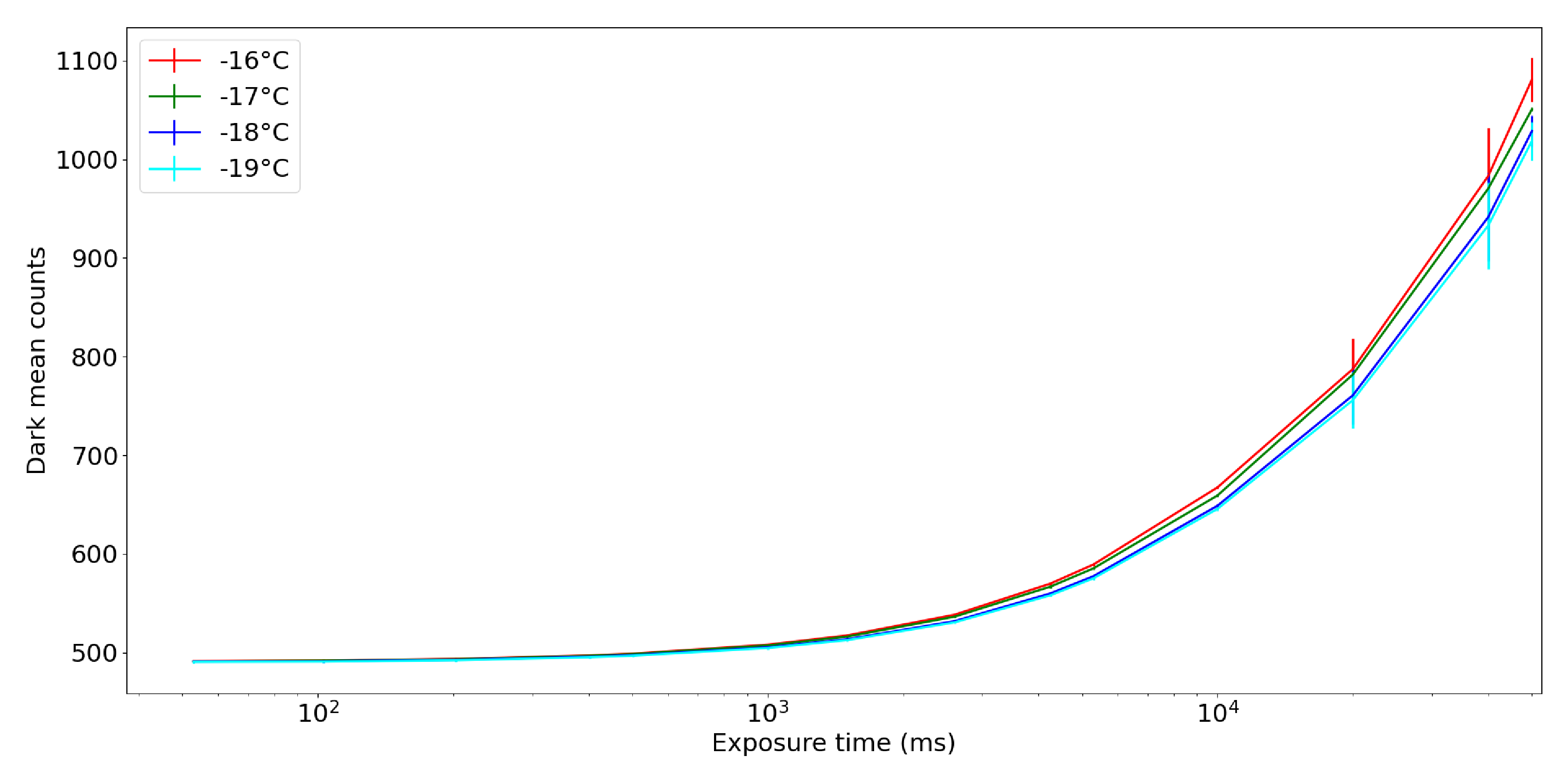}
        \caption{Mean dark variation for IR with deviation as error bar}
        \label{fig:dark40}
    \end{subfigure}
    
    \caption{Mean dark variation for sCMOS and IR detectors}\label{fig:dark_variation}
\end{figure}

\begin{figure}[!ht]
    \centering
    \begin{subfigure}{1.0\textwidth}
        \includegraphics[width=\textwidth]{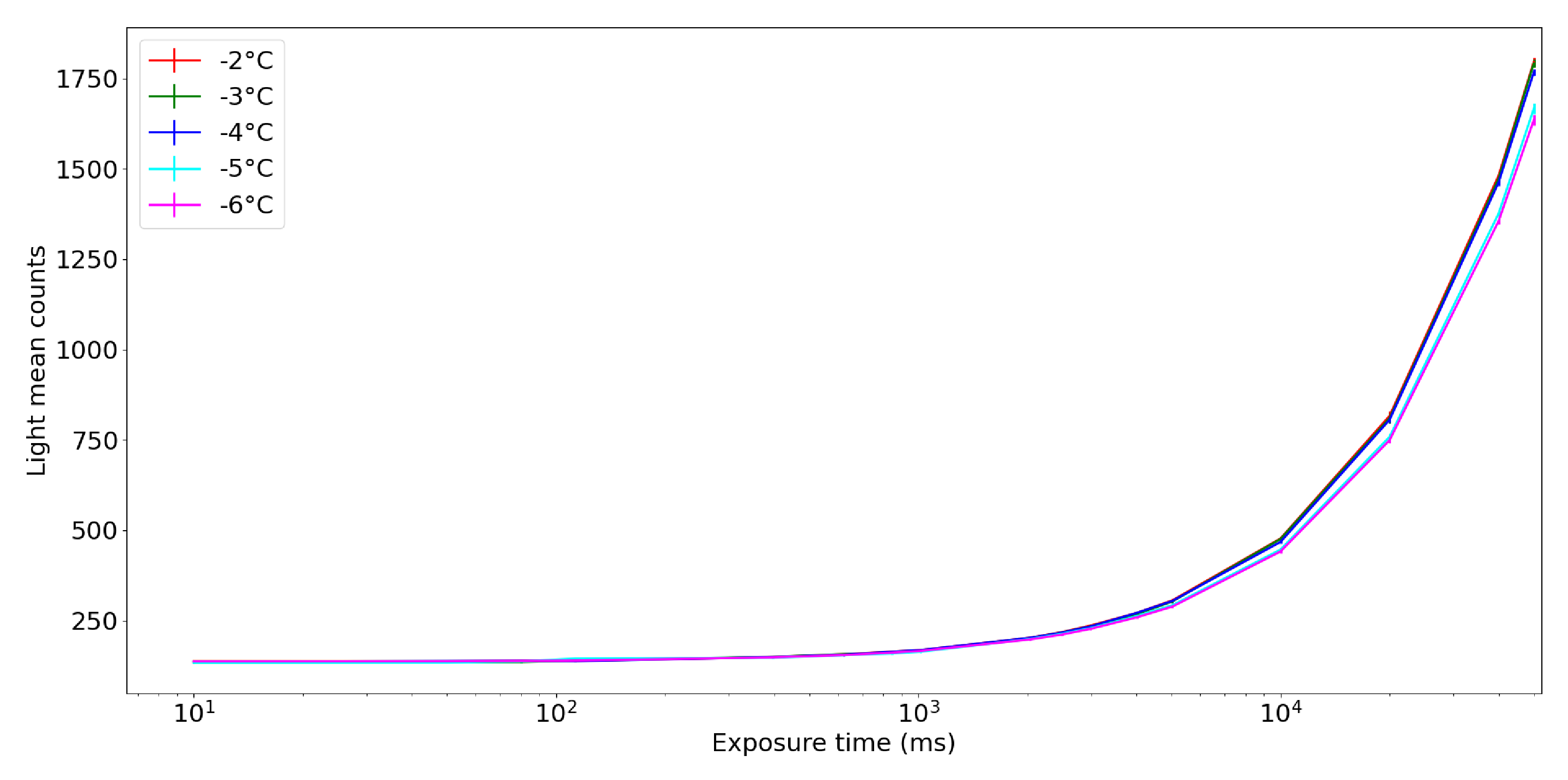}
        \caption{Light exposures for sCMOS with deviation as error bar}
        \label{fig:light30}
    \end{subfigure}
    
    \begin{subfigure}{1.0\textwidth}
        \includegraphics[width=\textwidth]{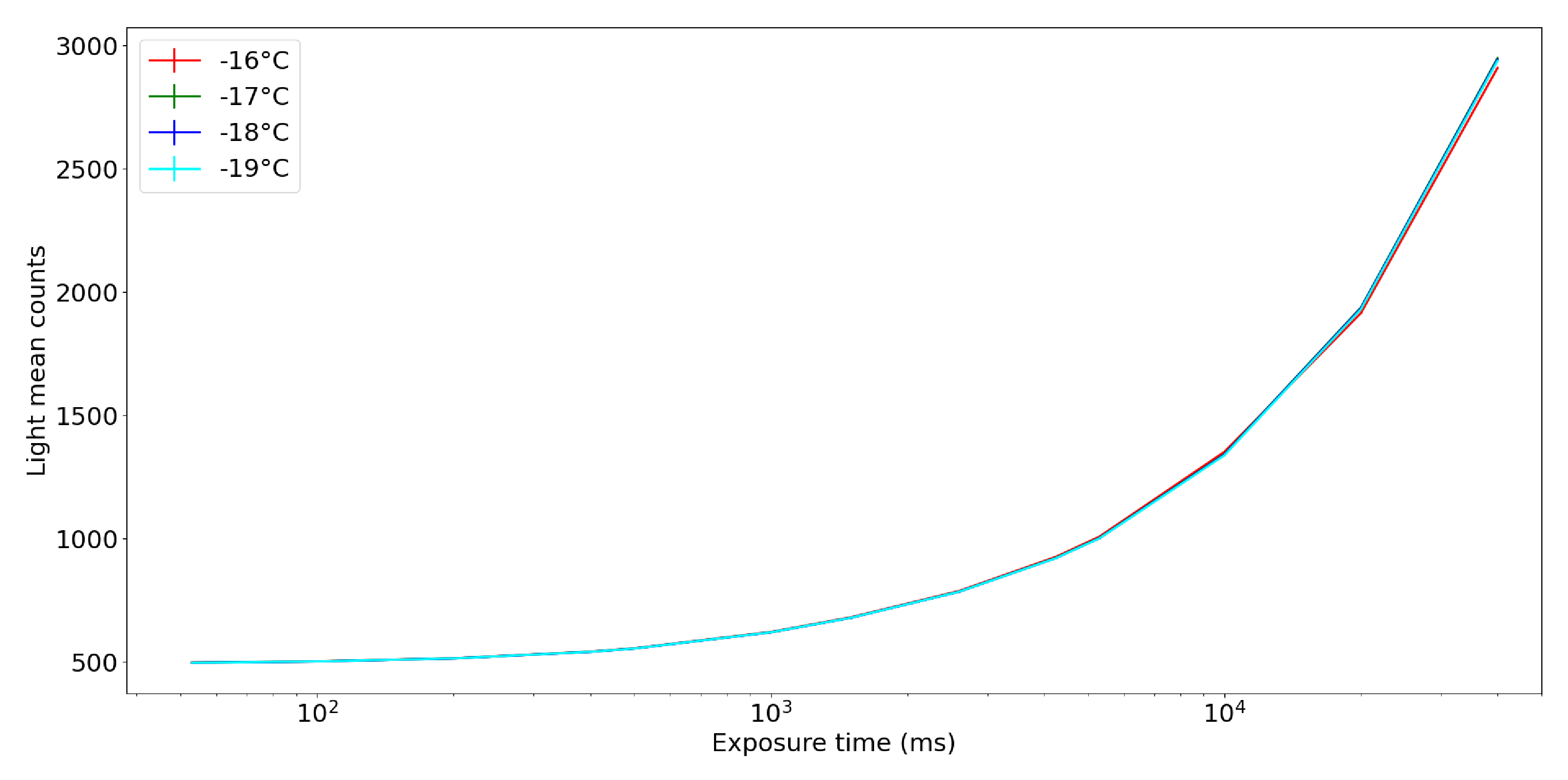}
        \caption{Light exposures for IR with deviation as error bar}
        \label{fig:light40}
    \end{subfigure}
    
    \caption{Light exposures for sCMOS and IR detectors}\label{fig:light_variation}
\end{figure}

The measured mean and standard deviation with respect to exposure time of acquired dark and light images for both visible and IR detectors at various temperatures is as shown in Figure \ref{fig:dark_variation} and Figure \ref{fig:light_variation}. 

\subsection{Estimation of detector parameters}

Using the light and dark data we estimated conversion gain, full-well capacity and read noise values using the standard method described in the following subsections \citep[for example refer][]{JANE2007, KAR2021,QIU2021,Gosain2022}.

\subsubsection{Conversion Gain}

The conversion gain is measured from the photon transfer curve obtained from the curve of variance with the mean photon counts during the light exposures of the detector. The curve is then fitted with line equation $y=mx+c$ where `m' and `c' are the slope and intercept of the curve. Furthermore, the conversion gain is calculated by taking the inverse of the obtained slope. The conversion gain of the visible detector is tabulated in Table \ref{CG30}, and the IR detector is tabulated in Table \ref{CG40}.

\begin{table}[!ht]
\begin{center}
\caption{Conversion Gain in $e^-/DN$ of visible detector}\label{CG30}%
\begin{tabular}{@{}llllllll@{}}
\toprule
S. No. & Gain  & $T=-2^\circ C$  & $T=-3^\circ C$  & $T=-4^\circ C$  & $T=-5^\circ C$  & $T=-6^\circ C$\\
\midrule
1.    & 1X   & 14.85 & 14.70 & 15.87 & 15.55 & 15.29\\
2.    & 2X   & 8.92  & 8.77  & 9.07  & 8.48  & 8.82\\
3.    & 10X  & 2.12  & 2.08  & 1.88  & 1.69  & 1.69\\
4.    & 30X  & 0.61  & 0.57  & 0.56  & 0.52  & 0.54\\

\botrule
\end{tabular}
\end{center}
\end{table}

\begin{table}[!ht]
\begin{center}
\caption{Conversion Gain in $e^-/DN$ of IR detector}\label{CG40}%
\begin{tabular}{@{}llllll@{}}
\toprule
S. No. & Gain  & $T=-16^\circ C$  & $T=-17^\circ C$  & $T=-18^\circ C$  & $T=-19^\circ C$ \\
\midrule
1.    & LG   & 768.13 & 790.65 & 779.62 & 781.84\\
2.    & HG   & 11.22  & 12.07  & 10.59  & 11.49 \\

\botrule
\end{tabular}
\end{center}
\end{table}

\subsubsection{Full Well Capacity}
The full-well capacity is obtained by multiplying the conversion gain with the mean counts when nearly 90\% of the saturation counts are achieved with light exposures. The full-well capacity of the visible detector is tabulated in Table \ref{FWC30}, and for IR detector is tabulated in Table \ref{FWC40}.

\begin{table}[!ht]
\begin{center}
\caption{Full well capacity in $ke^-$ of visible detector}\label{FWC30}%
\begin{tabular}{@{}llllllll@{}}
\toprule
S. No. & Gain  & $T=-2^\circ C$  & $T=-3^\circ C$  & $T=-4^\circ C$  & $T=-5^\circ C$  & $T=-6^\circ C$\\
\midrule
1.    & 1X   & 24.64 & 24.12 & 25.87 & 23.97 & 23.10\\
2.    & 2X   & 12.97  & 12.14  & 12.96  & 11.29  & 11.49\\
3.    & 10X  & 3.52  & 3.43  & 3.09  & 2.57  & 2.52\\
4.    & 30X  & 1.01  & 0.9  & 0.9  & 0.98  & 0.99\\

\botrule
\end{tabular}
\end{center}
\end{table}

\begin{table}[!ht]
\begin{center}
\caption{Full Well Capacity in $ke^-$ of IR detector}\label{FWC40}%
\begin{tabular}{@{}llllll@{}}
\toprule
S. No. & Gain  & $T=-16^\circ C$  & $T=-17^\circ C$  & $T=-18^\circ C$  & $T=-19^\circ C$ \\
\midrule
1.    & LG   & 1857.73 & 1943.72 & 1912.28 & 1913.82\\
2.    & HG   & 25.37  & 27.44  & 23.42  & 24.98 \\

\botrule
\end{tabular}
\end{center}
\end{table}

\subsubsection{Read Noise}
Read noise is obtained by multiplying the conversion gain with the standard deviation value for minimum possible exposure in the dark. The read noise of the visible detector is tabulated in Table \ref{RN30}, and the IR detector is tabulated in Table \ref{RN40}.

\begin{table}[!ht]
\begin{center}
\caption{Read noise in $e^-$ of visible detector}\label{RN30}%
\begin{tabular}{@{}llllllll@{}}
\toprule
S. No. & Gain  & $T=-2^\circ C$  & $T=-3^\circ C$  & $T=-4^\circ C$  & $T=-5^\circ C$  & $T=-6^\circ C$\\
\midrule
1.    & 1X   & 17.08 & 16.91 & 13.81 & 12.75 & 12.69\\
2.    & 2X   & 9.64  & 7.98  & 7.98  & 7.71  & 7.14\\
3.    & 10X  & 10.02  & 9.60  & 9.49  & 8.42  & 8.52\\
4.    & 30X  & 3.62  & 3.53  & 3.58  & 3.50  & 3.53\\

\botrule
\end{tabular}
\end{center}
\end{table}

\begin{table}[!ht]
\begin{center}
\caption{Read noise in $e^-$ of IR detector}\label{RN40}%
\begin{tabular}{@{}llllll@{}}
\toprule
S. No. & Gain  & $T=-16^\circ C$  & $T=-17^\circ C$  & $T=-18^\circ C$  & $T=-19^\circ C$ \\
\midrule
1.    & LG   & 988.33 & 956.15 & 942.31 & 865.0\\
2.    & HG   & 18.25  & 15.89  & 14.54  & 13.69 \\

\botrule
\end{tabular}
\end{center}
\end{table}

The Thermovac (TVAC) test of complete payload was carried out as part of environmental tests with complete thermal control in place. The dark counts of the detectors obtained during the standalone tests show a good correlation with the darks obtained during payload level (integrated VELC alone) and spacecraft level (after integration of the VELC and other payloads with the spacecraft).  Light exposures were taken only during the subsystem level tests and repetition of the same was not possible during the TVAC tests due to the contamination protocols. However, experiments related to payload calibration with solar disk light were carried out and the same was reported by Raghavendra Prasad et al \cite{BRP2023}.

VELC entrance aperture has a 45 mm neutral density filter with density 4 and that allows the part of the Sun light when pointed towards the Sun.  Therefore, it planned to point the spacecraft by 15-30 degrees away from the centre of the Sun and keep the EA door in closed condition and acquire the dark data. Similar dark measurements are planned to be carried out once in a three months for the entire mission lifetime. Such measurements help understand whether the onboard dark measurements are matching with the ones obtained during the ground based calibration and how the dark counts are varying with the mission life \cite{JS2022}. 

Further it is planned to carry out the disk images and disk spectral observations by pointing the spacecraft by 16 and 32 arcmin away from the disk centre and keeping the EA door in closed condition. When we off-point the spacecraft by 32 arcmin in four directions (say east, west, south and north) the disk of the Sun will appear in the field of view (FOV). When we off-point the spacecraft by 16 arcmin in four directions (say east, west, south and north), the centre of the Sun will fall in the slits and disk spectra can be obtained. Such observations will help understand the pixel response of the detector for the light \cite{JS2022}.

\section{Summary and Conclusions}

In this article, we discussed various technical details and specifications of the sCMOS and InGaAs detectors used in the Visible Emission Line Coronagraph. Furthermore, we presented the ground calibration methodology that opted to calibrate the detectors in the thermo-vacuum chamber located at the CREST campus of IIA. Using the observed parameters, we have estimated conversion gains, full-well capacity, readout noise, etc, for the detectors at various temperatures. Using such parameters, we concluded that operating the sCMOS and InGaAs detectors at $-5^{\circ}$ and $-17^{\circ}$ C onboard is essential.

\bmhead{Acknowledgments}
We thank all the Scientists/Engineers at the various centres of
ISRO such as URSC, SAC, LEOS, VSSC etc. and Indian Institute of Astrophysics who have made great contributions to the mission in order to achieve the present state. We thank anonymous referee for providing valuable suggestions that improved the manuscript.\\

\noindent {\bf Author's contributions}: SM and KSR analysed the data and wrote the manuscript. SM, KSR, SKVU, BH, UD involved in calibration activities and acquiring the data. BRP, JS, VSN, AK provided the inputs with the setup, designing calibration proceedures etc and also reviewed the manuscript. MP and SP helped with the data analysis. PUK, PK, NV, KS, VT, Suresha extended their help with the ETF facility, mounting the detectors in the TVAC chamber, maintaining the pressure and temperatures at desired levels etc. JHD, RK, SS, SK, ISB are the designers of the hardware electronics related to the detectors. \\

\noindent {\bf Funding} We gratefully acknowledge the financial support from Indian Space Research Organization (ISRO), India for this project.

\section*{Declarations}

{\bf Competing interests:} The authors declare no competing interests.\\ \\
{\bf Conficts of interest:} The authors declare that they have no confict of interest.

\bibliographystyle{unsrt}
\bibliography{main.bib}
\end{document}